\documentstyle[12pt]{article}
\newcommand{\be}[1]{\begin{equation}\label{#1}}
\newcommand{\ba}[1]{\begin{eqnarray}\label{#1}}
\newcommand{\ee}{\end{equation}}
\newcommand{\ea}{\end{eqnarray}}
\newcommand{\non}{\nonumber\\\rule{0pt}{30pt}}

\newcommand{\dis}{\displaystyle}
\newcommand{\eq}[1]{(\ref{#1})}
\newcommand{\tr}{\mathop{\rm tr}}
\renewcommand{\ll}[2]{{L}_{#1}(#2)}

\newlength{\HFPP}       \HFPP5.4mm
\makeatletter
\@addtoreset{equation}{section}
\makeatother

\newtheorem{thm}{Theorem}

\begin{document}
\begin{center}
\begin{Large}
{\bf Fredholm determinant representation for the partition
function of the six-vertex model}\\
\end{Large}
\vspace{20pt}
\begin{large}
N. A. Slavnov\\
\end{large}
\vspace{30pt}
{\it Steklov Mathematical Institute, Moscow, Russia.}\\
e-mail: nslavnov@mi.ras.ru\\
\vspace{32pt}
Abstract\\
\vspace{16pt}

\noindent\parbox{12cm}{\small
The six-vertex model with domain wall boundary conditions is considered.
A Fredholm determinant representation for the partition function of the
model is obtained. The kernel of the corrtesponding integral operator
depends on Laguerre polynomials.}

\end{center}

\vspace{32pt}
\section{Introduction}

In the present paper we consider the six-vertex model (ice model) on 
the two-dimensional square lattice with the special boundary 
conditions of ``domain wall" \cite{K}. First the six-vertex model was
solved in the papers \cite{L,S}, where the partition function and
bulk free energy were found for the case of periodically boundary 
conditions.  An explicit expression for the partition function 
with domain wall boundary conditions on the lattice of
finite volume  was found by A. G. 
Izergin \cite{I}. Izergin's representation contains a determinant of
$N\times N$ matrix (where $N^2$ is the number of vertices of the 
lattice). Thus, the problem of the thermodynamic limit of the 
partition function can be reduced to the study of the asymptotic 
behavior of the determinant at $N\to\infty$. 

The goal of this paper is to obtain new representation for the 
partition function, appropriate for the asymptotic analysis. A 
Fredholm determinant  representation seems to be suitable for
this goal. Let us explain in brief the main idea of the method 
suggested.

Let a matrix  $\cal M$ of the size $N$ is given, whose entries
${\cal M}_{jk}(t)$ depend on some parameter $t$. 
Determinants of this matrix $\det{\cal M}$  generate a
sequence of functions $f_N(t)$ for different $N$. 
As a rule the question on the properties of the limiting function 
$f_\infty(t)=\lim_{N\to\infty}f_N(t)$
is rather complicated. In spite of 
$\det{\cal M}$ is a finite sum of finite products, nevertheless
such ``explicit" expression for large $N$ becomes extremely 
complicated for analysis.

Suppose however that we can replace $\det{\cal M}$
with a Fredholm determinant
$$
\det{\cal M}=\det(I+V)
$$
of an integral operator, whose kernel parametrically depends on 
$t$ and $N$:  $V(x,y)=V_N(x,y|t)$. The Fredholm determinant is an infinite 
sum, and each term of the last one can be presented as a multiple 
integral, for example,
$$
\det(I+V)=\exp\left(\sum_{k=1}^{\infty}
\frac{(-1)^{k+1}}{k}\int
V_N(x_1,x_2|t)\cdots V_N(x_k,x_1|t)\,d^kx\right).
$$
While at finite $N$ the computations of the mentioned integrals are 
very complicated, at $N\to\infty$ the existence of large parameter in 
the integrands often allows to estimate not only each integral, but 
the series as a whole. Thus we have the possibility to obtain an 
estimate for the determinant. Hence, we can say that for $N\to\infty$ 
representations in terms determinants of finite size matrices become 
``less explicit", while Fredholm determinants turn to be ``more 
explicit".

Transformations 
of finite determinants into Fredholm ones were found 
to be useful, for instance, in asymptotic analysis of correlation 
functions of quantum integrable models, asymptotic behavior of 
orthogonal polynomials, the spectrum of random matrices etc. We hope 
that in the case under consideration the Fredholm determinant 
representation will permit eventually to solve the problem of the 
thermodynamic limit of the partition function of the six-vertex 
model\footnote{%
Recently the thermodynamic limit of the six-vertex model was studied
by use of Toda-chain equation \cite{KZ} and matrix models methods
\cite{Z}. The approach suggested in the present paper is completely 
different from the mentioned ones, and we hope that it is also worth
attention.}.

The content of the paper is as follows. In the next section the 
necessary information on the six-vertex model and Izergin's formula 
for the partition function are given. In the third section we 
transform  this formula to the Fredholm determinant representation.  
In the last section we discuss the outlook
for application of the obtained formula to the asymptotic analysis.

\section{The six-vertex model} 

Consider a statistical system on a square lattice of the size   
$N\times N$, associating with each edge some variables (classical 
spin), taking values $\pm1$.  Usually these variables are denoted by 
arrows, identifying $(\uparrow,~\rightarrow)$ with $1$, and 
$(\downarrow,~\leftarrow)$ with $-1$. Thus, in general case there 
are $16$ different configurations of arrows in each vertex. The 
six-vertex is specified by the condition that the number of arrows 
entering and leaving each vertex coincide. The types of possible 
vertices are shown on Fig. 1. With each configuration we associate a 
statistical weights. Generically the last ones can depend on the position 
of the vertex. We consider the model, which is symmetric with respect 
to simultaneous reversal of all arrows. Thus, we have $3N^2$ 
statistical weights $w_{jk}=\{a_{jk},~b_{jk},~c_{jk}\}$, where 
indices $j,k=0,1,\dots,N-1$ numerate vertices. The domain wall 
boundary conditions correspond to the entering arrows on the lower 
and upper boundaries, and leaving arrows on the left and right ones.

\begin{figure}[h] 
\begin{picture}(450,200)
\put(100,0){\begin{picture}(100,180)%
\put(0,30){\line(1,0){60}}
\put(0,30){\vector(1,0){15}}
\put(0,30){\vector(1,0){50}}
\put(30,0){\line(0,1){60}}
\put(30,0){\vector(0,1){15}}
\put(30,0){\vector(0,1){50}}
\put(0,130){\line(1,0){60}}
\put(60,130){\vector(-1,0){15}}
\put(60,130){\vector(-1,0){50}}
\put(30,100){\line(0,1){60}}
\put(30,160){\vector(0,-1){15}}
\put(30,160){\vector(0,-1){50}}
\put(30,180){$a_{jk}$}
\end{picture}}
\put(200,0){\begin{picture}(100,180)%
\put(0,30){\line(1,0){60}}
\put(60,30){\vector(-1,0){15}}
\put(60,30){\vector(-1,0){50}}
\put(30,0){\line(0,1){60}}
\put(30,0){\vector(0,1){15}}
\put(30,0){\vector(0,1){50}}
\put(0,130){\line(1,0){60}}
\put(0,130){\vector(1,0){15}}
\put(0,130){\vector(1,0){50}}
\put(30,100){\line(0,1){60}}
\put(30,160){\vector(0,-1){15}}
\put(30,160){\vector(0,-1){50}}
\put(30,180){$b_{jk}$}
\end{picture}}
\put(300,0){\begin{picture}(100,180)%
\put(0,30){\line(1,0){60}}
\put(60,30){\vector(-1,0){50}}
\put(0,30){\vector(1,0){50}}
\put(30,0){\line(0,1){60}}
\put(30,0){\vector(0,1){15}}
\put(30,60){\vector(0,-1){15}}
\put(0,130){\line(1,0){60}}
\put(0,130){\vector(1,0){15}}
\put(60,130){\vector(-1,0){15}}
\put(30,100){\line(0,1){60}}
\put(30,160){\vector(0,-1){50}}
\put(30,100){\vector(0,1){50}}
\put(30,180){$c_{jk}$}
\end{picture}}
\end{picture}
\caption{Verteces of the six-vertex model and their statistical
weights.}
\end{figure}
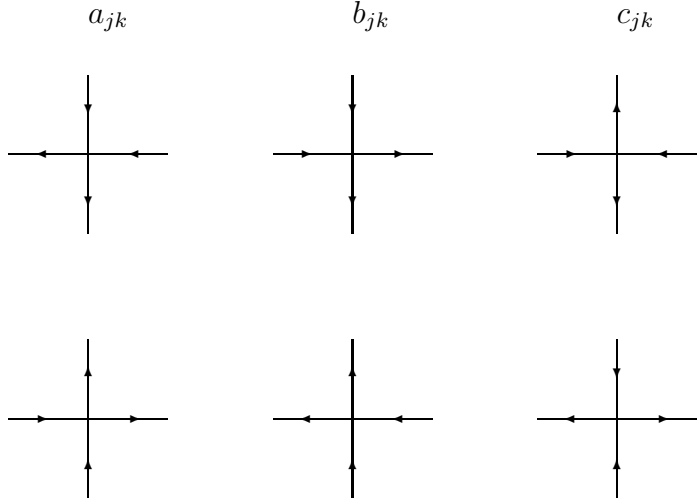
The partition function is given by
\be{parf}
Z_N=\sum_{\{C\}}\prod_{j,k=0}^{N-1}w_{jk}.
\ee
Here the sum is taken with respect to all possible configurations $C$
on the whole lattice.

The six-vertex model is exactly solvable, if $3N^2$ weights
$\{a_{jk},~b_{jk},~c_{jk}\}$ are parameterized by $2N+1$ variables
$\lambda_j$, $\xi_j$, $(j=0,1,\dots,N-1)$ and $\eta$:
\be{statvesa}
a_{jk}=\varphi(\lambda_j-\xi_k+\eta),\qquad
b_{jk}=\varphi(\lambda_j-\xi_k),\qquad
c_{jk}=\varphi(\eta).
\ee
Parameters $\lambda_j$ are associated with horizontal lines, 
$\xi_j$---with vertical ones (see. Fig. 2), $\eta$ is an arbitrary 
complex constant. Function  $\varphi(x)$ is defined up to 
normalization factor and is equal to
\be{2vozm} 
\varphi(x)=\sinh(x),\qquad\mbox{or}\qquad
\varphi(x)=x.
\ee
In the first case the model is equivalent to the quantum $XXZ$ magnet,
in the second---to the $XXX$ magnet.
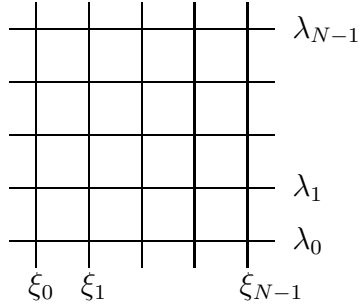
\begin{figure}[t]
\begin{picture}(400,180)%
\put(200,20){\begin{picture}(100,100)%
\multiput(20,0)(20,0){5}%
{\line(0,1){100}}
\multiput(10,10)(0,20){5}%
{\line(1,0){100}}
\end{picture}}
\put(180,0){\begin{picture}(100,100)%
\put(37,10){$\xi_0$}
\put(57,10){$\xi_1$}
\put(117,10){$\xi_{N-1}$}
\put(137,27){$\lambda_0$}
\put(137,47){$\lambda_1$}
\put(137,107){$\lambda_{N-1}$}
\end{picture}}
\end{picture}
\caption{Parametrization of the edges and verteces of the lattice.}
\end{figure}

The Izergin's formula for the partition function on the lattice of finite
volume with the domain wall boundary conditions has the form
\be{2parfun}
Z_N=\frac{\prod\limits_{a,b=0}^{N-1}
\varphi(\lambda_a-\xi_b+\eta)\varphi(\lambda_a-\xi_b)}
{\prod\limits_{N>a>b\ge0}
\varphi(\lambda_a-\lambda_b)\varphi(\xi_b-\xi_a)}
\det\left(\frac{\varphi(\eta)}{\varphi(\lambda_j-\xi_k)
\varphi(\lambda_j-\xi_k+\eta)}\right).
\ee
We would like to draw attention of the reader that here and further 
we numerate indices of matrices starting with zero, but not with one:  
$j,k=0,1,\dots,N-1$.

We are interesting in homogeneous limit, when the statistical 
weights do not depend on the position of the vertex. In other words
$\lambda_j=\lambda,\quad\xi_j=\xi,\quad
j=0,1\dots,N-1$. The corresponding limit in \eq{2parfun} can be 
obtained via
\be{homlim}
\lim_{\lambda_j\to\lambda\atop{\xi_k\to\xi}}
\frac{\det[\Phi(\lambda_j,\xi_k)]}{\Delta(\lambda)\Delta(\xi)}
=\det\left[\frac{1}{j!k!}
\frac{\partial^j}{\partial\lambda^j}
\frac{\partial^k}{\partial\xi^k}\Phi(\lambda,\xi)
\right].
\ee
Here $\Phi(\lambda,\xi)$ is some enough smooth two-variable function, 
$\Delta(\lambda)$ and 
$\Delta(\xi)$ are Van der Monde determinants of variables 
$\{\lambda\}$ and $\{\xi\}$ respectively.

In the rational case, when $\varphi(x)=x$  the direct use       
of \eq{homlim} gives
\be{partsumXXX}
Z_{N}^{(XXX)}=\frac{\Bigl[\nu(\nu+\eta)\Bigr]^{N^2}}
{\prod\limits_{m=0}^{N-1}(m!)^2}
\det\left[(j+k)!\left(\nu^{-j-k-1}-(\nu+\eta)^{-j-k-1}\right)\right],
\ee
where $\nu=\lambda-\xi$.

In the trigonometric case it convenient first to use variables
$u_j=e^{2\lambda_j}$, $v_j=e^{2\xi_j}$  and $q=e^\eta$.
Then \eq{2parfun} takes the form
\be{3parfun}
Z_N^{(XXZ)}=\prod_{a=0}^{N-1}\left(e^{\lambda_a-\xi_a}\right)
\frac{\prod\limits_{a,b=0}^{N-1}
\sinh(\lambda_a-\xi_b+\eta)(u_a-v_b)}
{\prod\limits_{N>a>b\ge0}
(u_a-u_b)(v_b-v_a)}
\det\left(\frac{1}{u_j-v_k}
-\frac{q}{qu_j-q^{-1}v_k}\right).
\ee
Now the equation \eq{homlim} can be easily applied,
and after simple algebra we obtain
\be{partsumXXZ}
Z_{N}^{(XXZ)}=e^{N\nu}\frac{
\Bigl[\sinh(\nu)\sinh(\nu+\eta)\Bigr]^{N^2}}
{\prod\limits_{m=0}^{N-1}(m!)^2}
\det\left[\frac{(j+k)!}
{\sinh^{j+k+1}(\nu)}-
\frac{q^{j-k+1}(j+k)!}
{\sinh^{j+k+1}(\nu+\eta)}\right].
\ee
Here as in \eq{partsumXXX} $\nu=\lambda-\xi$, and $q=e^\eta$.

Below we shall focus on the trigonometric case 
\eq{partsumXXZ}, as the most general one. As usual, the answer for 
the partition function $Z_{N}^{(XXX)}$ can be obtained in the limit
$\nu=\epsilon\nu,\quad \eta=\epsilon\eta,\quad \epsilon\to0$.

Our goal is to transform  \eq{partsumXXZ} to the expression, 
suitable for the asymptotic analysis at $N\to\infty$.  There are 
factors with trivial asymptotics in the equation \eq{partsumXXZ}: 
$e^{N\nu}$ and $\Bigl[\sinh(\nu)\sinh(\nu+\eta)\Bigr]^{N^2}$. 
It is also easy to evaluate the behavior of the factor
$\prod_{m=0}^{N-1}(m!)^2$ at  
$N\to\infty$.  The main problem is the determinant. In the next 
section we consider some special transforms of 
\eq{partsumXXZ}, which result finally to the new representation for 
the partition function, containing the determinant of an integral operator.

\section{Fredholm determinant}

As the first step, it is convenient to extract, for example, 
$\sinh^{-j-k-1}(\nu)$  out off  the 
determinant \eq{partsumXXZ}. Then \eq{partsumXXZ} becomes 
\be{start}
Z_{N}^{(XXZ)}=e^{N\nu}\frac{
\sinh^{N^2}(\nu+\eta)}
{\prod\limits_{m=0}^{N-1}(m!)^2}
\det[A-qtQTATQ^{-1}].
\ee

Here
\be{matrix}
A_{jk}=(j+k)!~,\qquad
T_{jk}=\delta_{jk}t^j,\qquad
Q_{jk}=\delta_{jk}q^j,\qquad
t=\frac{\sinh(\nu)}{\sinh(\nu+\eta)}.
\ee

{\sl Note.} Similarly we could extract terms 
$\sinh^{-j-k-1}(\nu+\eta)$. Then, up to the replacements           
$t\to t^{-1}$, $q\to q^{-1}$ and 
trivial common factor, the arising expression coincides with 
\eq{start}.

Now we can transform \eq{start} as follows:
\be{vydel}
\det[A-qtQTATQ^{-1}]=\det A\cdot\det[I-U],
\qquad\mbox{where}\qquad U=qtA^{-1}QTATQ^{-1}.
\ee
The advantage of this method is that $\det A$ 
cancels the product of factorials in the denominator, since
\be{detA} 
\det A=\prod_{m=0}^{N}(m!)^2,
\ee
(see Appendix 1). Thus our problem reduces to the derivation of 
the inverse matrix $A^{-1}$ and calculation of the product
$U=qtA^{-1}QTATQ^{-1}$. The inverse matrix $A^{-1}$ can be easily 
written in terms of Laguerre polynomials, therefore below we recall their
basic properties.

The Laguerre polynomials $\ll nx$ are defined by
\be{defL}
\ll{n}x=\frac{e^x}{n!} \frac{d^n}{dx^n}(x^ne^{-x})
\ee
and  generate the orthogonal system on the half-axis
$R_+$ with the weight $e^{-x}$:
\be{orthogL}
\int_0^\infty\ll nx\ll mx e^{-x}\,dx=\delta_{nm}.
\ee

The key object, determining the matrix $A^{-1}$, is the kernel
\be{defp}
K_n(x,y)=\sum_{k=0}^{n}\ll kx\ll ky.
\ee
One more useful representation for $K_n(x,y)$ follows from
Christoffel--Darboux formula:
\be{defp1}
K_n(x,y)=-\frac{n+1}{x-y}\Bigl(\ll {n+1}x\ll ny
-\ll {n+1}y\ll nx\Bigr).
\ee
Obviously, for arbitrary polynomial $\pi_m(x)$ of degree
$m$, less or equal to $n$, holds
\be{intpxm}
\int_0^\infty\pi_m(x) K_n(x,y)e^{-x}\,dx=\pi_m(y),
\qquad 0\le m\le n.
\ee
To prove \eq{intpxm}, it is enough to substitute to the l.h.s.
the expansion of $\pi_m(x)$ with respect to Laguerre polynomials
$\pi_m(x)=\sum_{k=0}^{m}c_k \ll kx$, and to use the definition
\eq{defp}.

Being a polynomial of variables $x$ and $y$, the kernel $K_n(x,y)$
can be presented in the form
\be{poljk}
K_n(x,y)=\sum_{j,k=0}^{n}K^{(n)}_{jk}x^jy^k,
\ee
where
\be{ochev}
K^{(n)}_{jk}=\left.\frac{1}{j!k!}
\frac{\partial^{j+k}}{\partial x^j\partial y^k}
K_{n}(x,y)\right|_{x=y=0}\quad.
\ee
Now we are in position to formulate the theorem on the inverse matrix.

\begin{thm}
Let $A_{jk}=(j+k)!$,\ $j,k=0,1\dots,N-1$. Then the inverse matrix
has the entries
\be{invA}
(A^{-1})_{jk}=K^{(N-1)}_{jk}.
\ee
\end{thm}

{\sl Proof.}~~Using the integral representation for
$(j+k)!$, we have
\be{interm2}
\sum_{l=0}^{N-1}A_{jl}K^{(N-1)}_{lk}=
\left.\frac{1}{k!}
\frac{\partial^k}{\partial y^k}\sum_{l=0}^{N-1}
\frac{1}{l!}\frac{\partial^{l}}{\partial x^l}
\int_0^\infty s^{j+l} e^{-s}K_{N-1}(x,y)\,ds
\right|_{x=y=0}\quad.
\ee
The sum with respect to $l$ is the Taylor series (in fact, since 
$K_{N-1}(x,y)$ is  $(N-1)$-degree polynomial of $x$, the 
mentioned series turns to the finite sum). Hence,
\be{interm3}
\sum_{l=0}^{N-1}A_{jl}K^{(N-1)}_{lk}=
\left.\frac{1}{k!}
\frac{\partial^k}{\partial y^k}
\int_0^\infty s^j e^{-s}
K_{N-1}(s,y)\,ds\right|_{y=0}\quad.
\ee
Due to \eq{intpxm} the integral in \eq{interm3} is equal to $y^j$.
Thus, we arrive at
\be{final}
\sum_{l=0}^{N-1}A_{jl}K^{(N-1)}_{lk}=
\left.\frac{1}{k!}
\frac{\partial^k}{\partial y^k}y^j\right|_{y=0}=\delta_{jk}.
\ee
The theorem is proved.

{\sl Note.}~~For generalization of the relationship between Hankel
matrices and Christoffel--Darboux kernels see \cite{Bor}.

The obtained explicit expression for $A^{-1}$ allows 
us to compute the matrix $U$ \eq{vydel}. Indeed,
\be{U1}
U_{jk}=\left.\frac{1}{k!}
\frac{\partial^k}{\partial y^k}\sum_{l=0}^{N-1}
t^{l+k+1}q^{l-k+1}
\frac{1}{l!}\frac{\partial^{l}}{\partial x^l}
\int_0^\infty s^{j+l} e^{-s}K_{N-1}(x,y)\,ds
\right|_{x=y=0}\quad.
\ee
Similarly to the proof of the Theorem the sum with respect to
$l$ easily can be computed and it gives
\be{resU}
U_{jk}=\frac{qt}{k!}\left.
\frac{\partial^k}{\partial y^k}
\int_0^\infty x^j(q^{-1}t)^k e^{-x}K_{N-1}(xqt,y)\,dx\right|_{y=0}.
\ee

Thus, the partition function $Z_N^{(XXZ)}$ turns to be proportional 
to the determinant of the matrix $\delta_{jk}-U_{jk}$ 
\eq{resU}.  To compute the last one it is convenient to use the
equation 
\be{spurlog} 
\det(I-U)=\exp\{\tr\log(I-U)\}= 
\exp\left\{-\sum_{n=1}^{\infty}\frac{1}{n}
\tr U^n\right\}.
\ee
Hence, we need to find the traces of powers of the matrix $U$.
It is easy to see that
\be{Usq}
(U^n)_{jk}=\left. \frac{(qt)^n}{k!}
\frac{\partial^k}{\partial y^k}
\int_0^\infty
x_1^j (q^{-1}t)^k K_{N-1}(x_1qt,x_2q^{-1}t)
\cdots K_{N-1}(x_nqt,y)
e^{-\sum_{m=1}^{n}x_m}\,d^nx
\right|_{y=0}.
\ee
Indeed, for $n=1$ \eq{Usq} coincides with \eq{resU}. Assuming that
\eq{Usq} holds for $U^n$, we immediately obtain
\ba{Un+1}
&&{\dis \sum_{l=0}^{N-1}(U^n)_{jl}U_{lk}= 
\frac{(qt)^{n+1}}{k!}
\frac{\partial^k}{\partial y^k}
\int_0^\infty
x_1^j(q^{-1}t)^k K_{N-1}(x_1qt,x_2q^{-1}t)\cdots }\non
&&{\dis\hspace{1cm}\left.
\cdots K_{N-1}(x_nqt,x_{n+1}q^{-1}t) K_{N-1}(x_{n+1}qt,y) 
e^{-\sum_{m=1}^{n+1}x_m}  \,d^{n+1}x
\right|_{y=0}.}
\ea
(Here the sum with respect to $l$ again turns to the Taylor series).
Hence,  
\be{trUn} 
\tr(U^n)= (qt)^n\int_0^\infty
K_{N-1}(x_1qt,x_2q^{-1}t)
\cdots K_{N-1}(x_nqt,x_1q^{-1}t)
e^{-\sum_{m=1}^{n}x_m}  \,d^nx.
\ee

The obtained expression \eq{trUn} for the trace of $U^n$ allows to
replace $\det(\delta_{jk}-U_{jk})$ with Fredholm determinant. In 
order to do this, we introduce integral operator $I-V$, acting on 
$R_+$ 
\be{actV}
[(I-V)f](x)=f(x)-\int_0^\infty V(x,y)f(y)\,dy,
\ee
with the kernel
\be{kern}
V(x,y)=qtK_{N-1}(xqt,yq^{-1}t)e^{-(x+y)/2}
=qt\sum_{k=0}^{N-1}\ll k{xqt}\ll k{yq^{-1}t}e^{-(x+y)/2}.
\ee
Due to \eq{defp1}, $V(x,y)$ also can be presented in the form
\be{kern1}
V(x,y)=
\frac{qN}{qx-q^{-1}y}
\left(\ll {N-1}{xqt}\ll N{yq^{-1}t}-
\ll N{xqt}\ll {N-1}{yq^{-1}t}\right)e^{-(x+y)/2}.
\ee

Then due to \eq{trUn}
\be{trtr}
\tr U^n = \tr V^n,
\ee
and  we finally obtain
\be{freddet}
\det(\delta_{jk}-U_{jk})=\det(I-V).
\ee

\section{Discussions}

Thus, we have found new representation for the partition function 
of the six-vertex model, containing the Fredholm determinant of the 
integral operator with the kernel \eq{kern}, \eq{kern1}:
\be{resXXZ}
Z_{N}^{(XXZ)}=e^{N\nu}
\Bigl[\sinh(\nu+\eta)\Bigr]^{N^2}\det(I-V).
\ee
Since the kernel $V$ is degenerated, we always can turn back to the 
determinant of the finite size matrix
\be{konpor}
\det\left(I-qt\sum_{k=0}^{N-1}\ll k{xqt}\ll k{yq^{-1}t}e^{-(x+y)/2}
\right)=
\det\left(\delta_{jk}-qt\int_0^\infty\ll j{xqt}
\ll k{xq^{-1}t}e^{-x}\,dx \right).
\ee
However, for the asymptotic analysis at $N\to\infty$ the 
Fredholm determinant representation seems to be preferable due to 
several reasons.

First, the equation \eq{intpxm} shows that the kernel $K_N(x,y)$ acts 
as identity operator on the subspace of the polynomials of $N$-th 
degree. Then one can think that for $N$ going to infinity, 
$K_N(x,y)$ goes to identity operator (delta-function). 
Of course, this limiting procedure requires a serious analysis of 
convergency. It is remarkable however, that for certain values of $q$ 
and $t$ such a naive interpretation of the kernel \eq{kern} does 
holds, and the Fredholm determinant can be evaluated explicitly in the
limit $N\to\infty$.

Second, similar integral operators with the kernels, depending on 
Laguerre polynomials, often appear in the theories of random
matrices and random permutations \cite{Met,Tr1}.  Some methods of the 
analysis of such operators can be directly applied to the case under 
consideration. Besides, at large $N$ the Laguerre polynomials
can be approximated by Bessel functions, what also leads us to the 
integral operator, analogous to ones, considered in the papers 
\cite{Nak,Tr2,Tr3}.

Third, in particular case of the partition function
$Z_{N}^{(XXX)}$, when $q=1$, the kernel $V(x,y)$ takes the form
\be{kernXXX}
V(x,y)=
\frac{N}{x-y}
\left(\ll {N-1}{xt}\ll N{yt}-
\ll N{xt}\ll {N-1}{yt}\right)e^{-(x+y)/2}.
\ee
The integral operator with the kernel \eq{kernXXX} belongs to the 
class of integrable integral operators \cite{IIKS}. 
The corresponding Fredholm determinant turn to be 
$\tau$-function of classical exactly solvable equation and can be
evaluated via the matrix Riemann--Hilbert problem methods. 

We are planning to present the detailed 
asymptotic analysis of the partition function of the six-vertex model 
with domain wall boundary conditions in forthcoming publications.

\vspace{5mm}\par

The author thanks A. R. Its for numerous and useful discussions. The 
work was supported in parts by RFBR, Grant 99-01-00151 and
INTAS-99-1782.

\appendix
\section{Calculation of the determinant}

One of the method to proof \eq{detA} is to use well known formula for
Cauchy determinant
\be{Cauchy}
\frac
{\prod\limits_{N>a>b\ge0}(\alpha_a-\alpha_b)(\beta_b-\beta_a)}
{\prod\limits_{a,b=0}^{N-1}(\alpha_a-\beta_b)}
=\det\left(\frac{1}{\alpha_j-\beta_k}
\right).
\ee
Here $\{\alpha\}$ and $\{\beta\}$ are arbitrary complex. Dividing both
sides of \eq{Cauchy} by Van der Monde determinants
$\Delta(\alpha)$ and $\Delta(\beta)$ and taking homogeneous limit
$\alpha_j\to\alpha,\quad\beta_j\to\beta,\quad j=0,1,\dots,N-1$, we  
obtain due to \eq{homlim}
\be{Cauchy1}
(\alpha-\beta)^{-N^2}=\prod_{m=0}^{N-1}(m!)^{-2}
\det\left(\frac{(j+k)!}{(\alpha-\beta)^{j+k+1}}\right),
\ee
what immediately implies \eq{detA}.

Another method to compute $\det A$ is based on the application of 
Laguerre polynomials. Below we present this method, since it allows
to find $A^{-1}$ as well as to prove \eq{detA}.

Using the integral representation for factorial, we have
\be{intG}
\det A=\det\left(\int_0^\infty\, e^{-s} s^{j+k}\,ds\right).
\ee
Denote integration variable in the
$j$-th row as $s_j$. Then we come to the multiple integral
\be{multint}
\det A=\int_0^\infty\,ds_0\cdots ds_{N-1}
\det\left( e^{-s_j}s^{j+k}_j\right)=
\int_0^\infty\,ds_0\cdots ds_{N-1}
\prod_{j=0}^{N-1}\left( e^{-s_j}s^{j}\right)\Delta(s),
\ee
where $\Delta(s)$ is Van der Monde determinant. Now we can use 
well known identity
\be{ident}
\Delta(s)=\det[p_k(s_j)],
\ee
where $p_k(s)$ is a system of {\sl arbitrary} polynomials with the highest
coefficient equal to $1$:
\be{pk}
p_k(s)=s^k+o(s).
\ee
In particular we can choose the Laguerre polynomials 
\be{vybor}
p_k(s)\equiv \tilde L_k(s)=
(-1)^ke^s\frac{d^k}{ds^k}(s^k e^{-s}).
\ee
The normalization in \eq{vybor}  differs from one in
\eq{defL} by factor $(-1)^k k!$ in order to provide \eq{pk}.
Therefore we denote the
polynomials \eq{vybor} as $\tilde L_k$. The orthogonality condition then 
takes the form
\be{orthonew}
\int_0^\infty\tilde L_m(s)\tilde L_n(s)e^{-s}\,ds=(n!)^2\delta_{nm}.
\ee
Thus, the equation \eq{multint} becomes
\be{multint1}
\det A=
\int_0^\infty\,ds_0\cdots ds_{N-1}
\prod_{j=0}^{N-1}\left( e^{-s_j}s^{j}\right)\det[
\tilde L_k(s_j)].
\ee
Integrating inside the determinant, we obtain
\be{tt}
\det A=\det\left(\int_0^\infty\, e^{-s} s^{j}
\tilde L_k(s)\,ds\right).
\ee
Due to orthogonality of the polynomials
$\tilde L_k(s)$, the entries in
\eq{tt} vanish, if $j<k$. Thus we deal with the determinant of 
triangular matrix and, hence,
\be{ttt}
\det A=\prod_{m=0}^{N-1}\left(\int_0^\infty\, e^{-s} s^{m}
\tilde L_m(s)\,ds\right).
\ee
It remains to observe that expansion  
of $s^m$ with respect to $\tilde L_m(s)$ has the form
$$
s^m=\tilde L_m(s)\quad+\quad\mbox{polynomials of lower degree}
$$
therefore due to \eq{orthonew}
\be{ans}
\det A=\prod_{m=0}^{N-1}(m!)^2,
\ee
which ends the proof.

The described method allows to compute not only $\det A$, but the 
minors of the size $N-1$ as well. This provides the possibility to
find $A^{-1}$.


\begin{thebibliography}{99}
%
\bibitem{K} V.~E.~Korepin,
Commun. Math. Phys. {\bf 86} (1982)  391.
%
\bibitem{L} E. H. Lieb, Phys. Rev.
{\bf 18} (1967) 1046;
{\bf 19} (1967) 108
%
\bibitem{S} B. Sutherland,  Phys. Rev. Lett.
{\bf 19} (1967) 103;
%
\bibitem{I} A. G. Izergin, 
Sov. Phys. Dokl. {\bf 32} (1987) 878
%
\bibitem{KZ} V. E. Korepin and P. Zinn-Justin, preprint 
cond-mat/0004250
%
\bibitem{Z} P. Zinn-Justin, preprint math-ph/0005008
%
\bibitem{Bor} A. Borodin, preprint math.CA/9804027
%
\bibitem{Met} M. L. Mehta, {\it Random Matrices},
2nd ed., Academic Press, San Diego, 1991
%
\bibitem{Tr1} C. A. Tracy and H. Widom, preprint math.CO/9904042
%
\bibitem{Nak} S. Nishigaki, Phys. Lett.
{\bf B387} (1996) 139
%
\bibitem{Tr2} C. A. Tracy and H. Widom,
Commun. Math. Phys., {\bf 161} (1994) 289
%
\bibitem{Tr3} E. L. Basor and C. A. Tracy,
J. Stat. Phys., {\bf 73} (1993) 415
%
\bibitem{IIKS}A. R. Its, A. G. Izergin, V. E. Korepin,                  
N. A. Slavnov,
Int. Journ. Mod. Phys.   {\bf B4} (1990) 1003.  
%
\end{thebibliography}
\end{document}